\documentclass[prb,preprint,showpacs,superscriptaddress,floatfix]{revtex4-1}
\usepackage{amssymb,mathptm}
\usepackage{amsmath,amssymb,amsfonts,bm}
\usepackage{graphicx,epsfig,color}

\begin{document}
\setcounter{page}{1}
\title{Reduction of charge fluctuation energies in ultrathin NiO films on Ag(001)}
\author{Seolun \surname{Yang}}
\affiliation{Department of Physics, Sookmyung Women's University, Seoul
140-742, Korea}
\author{H.-K. \surname{Park}}
\affiliation{Department of Physics, Sookmyung Women's University, Seoul
140-742, Korea}\emph{}
\author{J.-S. \surname{Kim}}
\affiliation{Department of Physics, Sookmyung Women's University, Seoul
140-742, Korea}
\author{H. -N. \surname{Hwang}}
\affiliation{Pohang Accelerator Laboratory, Pohang 790-784, Korea}
\author{C.-C. \surname{Hwang}}
\affiliation{Pohang Accelerator Laboratory, Pohang 790-784, Korea}
\author{S.-H. \surname{Phark}}
\affiliation{Department of Physics, Seoul National University, Seoul 151-747, Korea }
\author{Young Jun \surname{Chang}}
\affiliation{Department of Physics, Seoul National University, Seoul 151-747, Korea }
\author{T. W. \surname{Noh}}
\affiliation{Department of Physics, Seoul National University, Seoul 151-747, Korea }
\author{H.-D. \surname{Kim}}
\affiliation{Pohang Accelerator Laboratory, Pohang 790-784, Korea}

\email{hdkim@pal.postech.ac.kr, jskim@sm.ac.kr}
\date{\today}

\begin{abstract}
We investigate the on-site Coulomb interaction energy $U$(Ni $3p$) between two $3p$ holes in an ultrathin NiO film on Ag(001)
using both x-ray photoelectron spectroscopy and Auger electron spectroscopy.
As the film becomes thinner, $U$(Ni $3p$) decreases monotonically, and
the difference between the values of $U$(Ni $3p$) for a 1-monolayer (ML) film and a bulk-like thick film, $\delta U$(Ni $3p$), reaches $\sim - 2.2$ eV.
The observed value of $\delta U$(Ni $3p$) for a 1 ML film is well reproduced by the differences between both the image potentials and polarization energies
of the 1 ML film and the bulk-like thick film.
Hence, the present results provide evidence in favor of the reduction of charge fluctuation energies in ultrathin films on highly polarizable substrates, as
originally predicted by Duffy {\it et al.} [J. Phys. C: Solid State Phys., {\bf 16}, 4087 (1983)] and Altieri {\it et al.}[Phys. Rev. B {\bf 59}, R2517 (1999)].

\end{abstract}



\maketitle

\section{Introduction}

The electronic properties of ultrathin films are significantly different from those of their bulk states due to their reduced dimensionality
and the influence of substrates. In the case of oxides on metallic substrates, it has been predicted that characteristic 
features such as the on-site Coulomb interaction energy $U$ and the charge transfer energy $\Delta$ from a ligand
to a neighboring cation would be substantially altered for
atomically thin films.\cite{Duffy,Altieri}
Altieri {\it et al.}\cite{Altieri} observed that for an ultrathin MgO film on Ag(001),
both $U$ and $\Delta$ decreased monotonically as the film became thinner.
They attributed the decrements from the bulk values of $U$ and $\Delta$ in the film, $\delta U$ and $\delta \Delta$, to the extra-atomic relaxation energies $E_{rlx}$
that developed in response to the altered charge states of the ions. As the major sources of $E_{rlx}$, they considered both the image charge potential
energy $E_{image}$ between an extra charge and its image induced in the metal substrate and the polarization
energy $E_{pol}$ of the oxide caused by the extra charge also known as the Madelung potential energy. The magnitude of the image potential energy should be larger for 
thinner films due to the smaller mean distance between an extra charge in the film and its image in the substrate. The polarization energy
of the film should be different from that of the bulk oxide because the volume of the oxide in the film is reduced, whereas the polarizability
is enhanced at the surface of the oxide. The resulting variation  between $E_{rlx}$ in the film and in the bulk state, $\delta E_{rlx}$, even quantitatively reproduced
 the experimental $\delta U$ for a 1-monolayer (ML) MgO film on Ag(001).\cite{Altieri}

Nevertheless, the suspicion was raised that the successful reproduction of the experimental $\delta U$ by $\delta E_{rlx}$
could have been fortuitous, as there were many other effects (such as dipole-dipole interactions) that were not taken into account
as well as unjustified assumptions (such as $1/r$ dependence of the image potential on the atomic distance).
Moreover, Chambers and Droubay\cite{Chambers} reported that both Fe$_{2}$O$_{3}$ and Cr$_{2}$O$_{3}$ films on Pt(111) exhibited
negligible $\delta U$ and $\delta \Delta$.
This contrasting observation was attributed to effective intrinsic screening of charge transfer, which reduced
 the extra-atomic relaxation to an undetectable level.
Thus, no comprehensive elucidation of the electronic properties of ultrathin oxide films on highly polarizable substrates 
seems to exist, 
and the number of experimental studies are very limited to properly assess the existing hypotheses.

The objective of the present work is to assess the existing hypotheses by comparing their predictions with experimental results for a different
system: ultrathin NiO films on Ag(001). Bulk NiO is prototypical as a charge-transfer insulator, \cite{Zaanen} and its
 charge fluctuation energies have already been studied.\cite{Guenseop, Taguchi} Furthermore, the lattice mismatch between
NiO(001) and Ag(001) is only $\sim$ 2 $\%$, and the pseudomorphic growth of an NiO film is well established.\cite{Neddermeyer,Wollschlager,Caffio}
In other words, NiO films grown on Ag(001) are well suited for studying the thickness dependency of charge fluctuation energies such as $U$ and $\Delta$.

However, it is difficult to obtain the Coulomb interaction energy $U$(Ni $3d$) between Ni $3d$ electrons 
via the method of Altieri {\it et al.},\cite{Altieri}
because the Ni $3d$ spectrum is difficult to isolate due to its overlap with the Ag $4d$ band of the substrate.
Instead, we study the interaction energy between Ni $3p$ holes $U$(Ni $3p$).
As the film becomes atomically thin, $U$(Ni $3p$) exhibits a substantial reduction from its bulk value.
Moreover, the extra-atomic relaxation energies represented by both $E_{image}$ and $E_{pol}$
well reproduce the change in $U$(Ni $3p$) from bulk to thin film, $\delta U$(Ni $3p)$, for a 1 ML  NiO film on Ag(001).
Using the observed values of $\delta U$ and $\delta \Delta$, we estimate the N\'{e}el temperature $T_{N}$
 in the mean field approximation that is found to be compatible with the experimental value of $T_{N}$ for a 3-ML NiO film.\cite{Tjeng}
These results reinforce the idea that the extra-atomic relaxation represented mainly
 by $E_{image}$ and $E_{pol}$ determines $\delta U$ and $\delta \Delta$
 for ultrathin oxide films of NiO, as well as MgO, on highly polarizable substrates.

\section{Experiment}

We performed {\it in situ} scanning tunneling microscopy (STM), photoelectron spectroscopy (PES), and Auger electron spectroscopy (AES) on ultrathin NiO films grown on Ag(001).
The STM work was carried out with a variable-temperature STM (Omicron).
The NiO films were grown in an attached preparation chamber, where the preliminary characterization of both Ag substrate
and NiO film was accomplished by x-ray photoelectron spectroscopy (XPS) and low-energy electron diffraction (LEED).

The PES and AES work were carried out with a soft x-ray beamline (7B1) at Pohang Light Source in Korea.
The end station of the beamline is composed of both an analysis chamber and a preparation chamber.
The analysis chamber is equipped with a hemispherical electron energy analyzer with a multichannel detector.
For the PES, the photoelectrons are collected at a take-off angle of 45$^{\circ}$ with respect to the surface normal of the sample.
The PES resolving power is $\sim$ 4000.\cite{HN}
The zero point of the binding energy is determined in reference to the binding energy of the Ag $3d$ (368.3 eV) of the clean Ag substrate.
All spectra presented in this work were recorded with the sample maintained at room temperature.
For both STM and PES, no charging effects were observed.

The NiO films were grown in preparation chambers for both STM and PES. The base pressures were
$<$ $5\times 10^{-10}$ Torr for both chambers. Wedge-shaped NiO films were grown by e-beam evaporation of high purity (5N) Ni
rod onto clean Ag(001) at room temperature at an ambient O$_{2}$ pressure ($P_{{\rm O}_{2}}$) of $1 \sim 3 \times 10^{-6}$ Torr.
The films were then thermally annealed at $430 \sim 450$ K at $P_{{\rm O}_{2}}$ 
$\sim 5 \times 10^{-7}$ Torr.
In the present work, we are especially interested in films within the monolayer limit,
as this enables a definite comparison of experimental $\delta U$ with theoretical values obtained by considering extra-atomic relaxation energies.
However, for films less than 2 ML, the growth mode is somewhat complicated due to the
($2 \times 1$) reconstruction and the bilayer growth of the NiO film.\cite{Neddermeyer, Wollschlager, Caffio}
Under the aforementioned growth conditions, we were able to grow 1 ML ($1 \times 1$) nickel oxide films, as assessed by a combination of techniques,
including STM, LEED, and XPS. (Further details are given in the following section.)
According to our previous extensive PES of NiO films, such growth conditions also minimize the chemical defects.\cite{chemical}

The thickness of each film was mainly determined by the ratio of the peak intensity of the Ag $3d$ in the NiO-covered region
to that of the clean Ag substrate, assuming layer-by-layer (LBL) growth of the film.
Because the growth of a NiO film does not follow LBL in an ideal fashion, the thicknesses described in the present work are nominal.
For coverage $\sim$ 0.5 ML, the film is mainly composed of monolayer-high islands and can be taken as a model for a 1 ML film (Fig. 1(a)). 
(Further discussion is presented below.)
Moreover, up to 0.5 ML, the coverage recorded by a quartz microbalance is in reasonable agreement
with the nominal coverage estimated by the reduction of Ag $3d$ intensity, assuming layer-by-layer growth of the NiO film.
Based on these estimated film thicknesses, the growth rate is adjusted to $\sim$ 0.25 ML/min throughout the experiments.

\section{Result}

\begin{figure}
\centering
\includegraphics[width=0.75\textwidth]{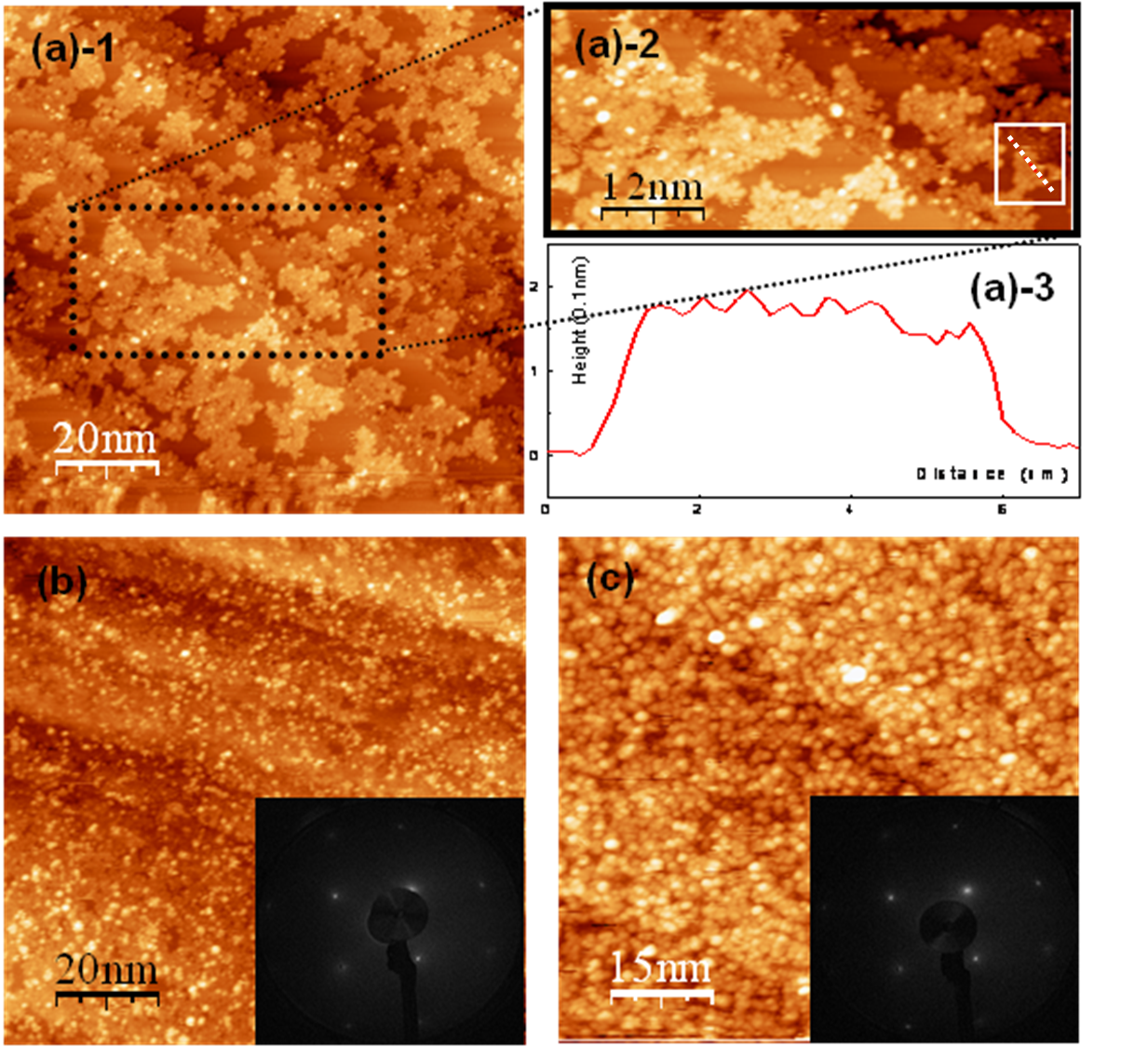}
\caption[] {NiO films grown on Ag(001) at room temperature via e-beam evaporation of Ni at a $P_{O_{2}}$ of $3 \times 10^{-6}$ Torr.
The NiO film coverage is (a) 0.5 ML, (b) 0.67 ML, and (c) 1 ML. The scanning voltage and tunneling current were
-2.7 V (sample) and 0.1 nA, respectively. Figs. (b) and (c) show LEED images of ($1 \times 1$) patterns at an electron energy of 137 eV.}
\label{fig.1}
\end{figure}

Figure 1 (a)-1 shows a typical image of a nickel oxide film with $\sim$ 0.5 ML coverage, consisting of nickel oxide patches.
The line profile (Fig. 1 (a)-3) across a typical patch in Fig. 1 (a)-2 displays a plateau of apparent height $\sim 0.15$ nm, which corresponded to 1 ML in our previous STM study of NiO film on Ag(001) under similar tunneling conditions.\cite{STM} 
As the film (with its nominal coverage of $\sim$ 0.5 ML) is mostly composed of islands of thickness 1 ML, 
we regard its Ni and O spectra asrepresentative of the electronic properties of a 1 ML NiO film.
After further deposition to the nominal coverage is around 1 ML, the second layer is preferably occupied (Fig. 1 (b)), and the film is almost bilayered. (Fig. 1 (c)) All films exhibit the ($1 \times 1$) LEED pattern (Fig. 1 (b) and (c)), whereas the well-known ($2 \times 1$) reconstruction is observed only sporadically in the STM images (Figure not shown). However, the ($2 \times 1$) reconstruction becomes abundant if $P_{{\rm O}_{2}}$ is lowered beneath 10$^{-6}$ Torr.

\begin{figure}
\includegraphics[width=1\textwidth]{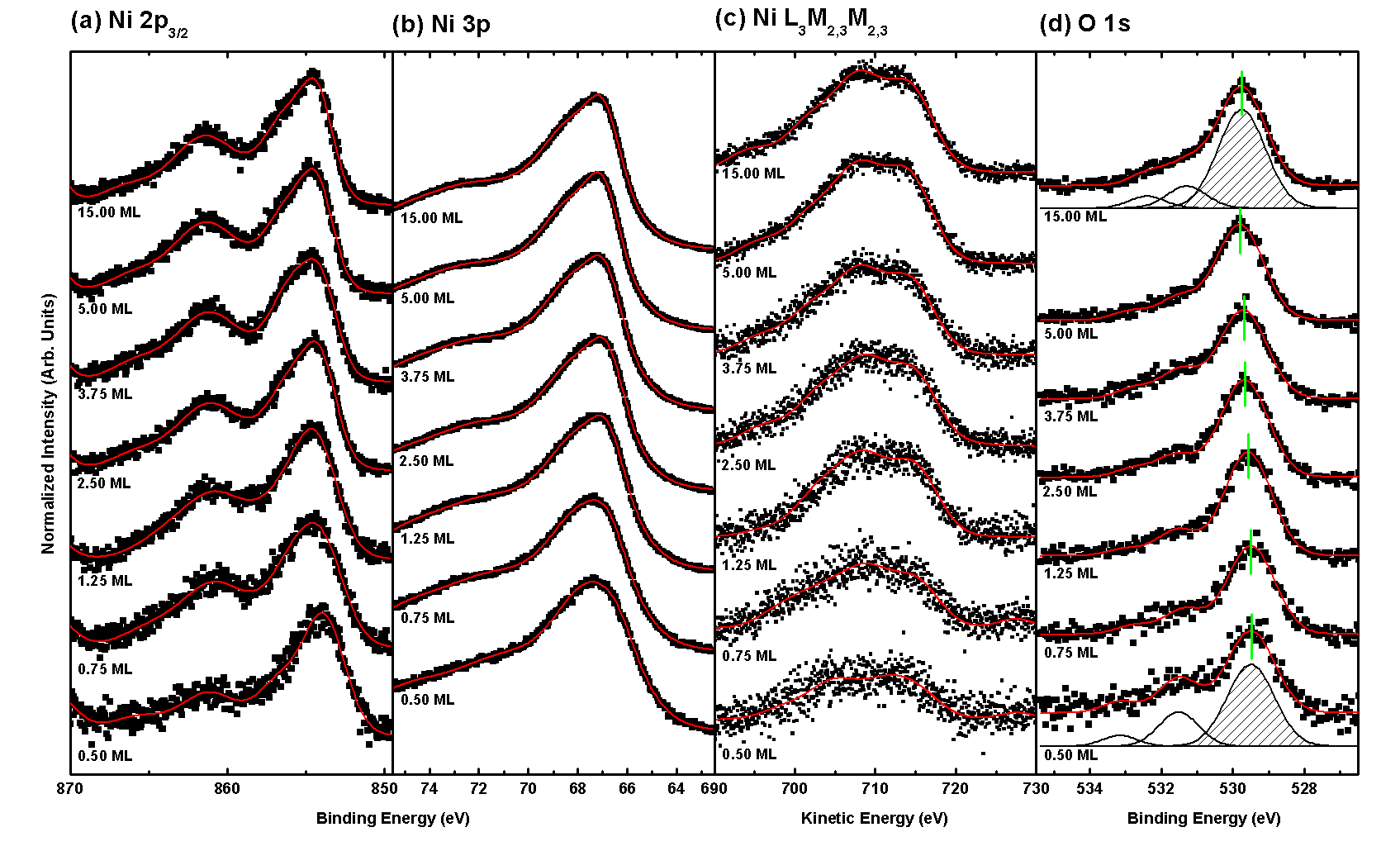}
\caption[width=0.5\textwidth] {As the ultrathin nickel oxide film on the Ag(001) substrate becomes thinner, the centroids of the (a) Ni $2p_{3/2}$ and (b) Ni $3p$ spectra shift to the lower binding energy side, while (c) the centroid of the Ni$_{LMM}$ Auger transition moves to the higher kinetic energy side. (d) The binding energy of the main peaks of the O $1s$ spectra becomes smaller as the film becomes thinner. The film thickness ranges from 0.5 to 15.00 ML. All the spectra are normalized by the incident photon intensity.}
\label{fig.2}
\end{figure}

$U$(Ni $3p$) can be obtained by comparing the energy of a two-hole state, Ni $3p^{4}$, to that of two one-hole state, Ni $3p^{5}$,
in accordance with the relationship

\begin{equation}
U(Ni\;\;3p) = E(3p^{4}) + E(3p^{6}) - 2E(3p^{5}).
\label{eqn:1}
\end{equation}
The variation of $U$(Ni $3p$) from that of bulk (actually bulk-like thick film) $\delta U$(Ni $3p$) can then be determined by the following relation,

\begin{equation}
 \delta U(Ni\;\;3p) = \delta E^{bind}(Ni\;\;2p) - 2 \delta E^{bind}(Ni\;\;3p) - \delta E^{kin}(Ni_{LMM}),
\label{eqn:2}
\end{equation}
which is obtained via the approach of Altieri {\it et al.}\cite{Altieri}
To estimate $\delta U$(Ni $3p$), we measured the XPS spectra of Ni $2p$, Ni $3p$, and the Ni $LMM$ Auger transition as functions of the thickness of NiO film. We also utilized the O 1s spectra to estimate $\delta \Delta$(O $2p$ $\rightarrow$ Ni $3d$), as described below.

Figure 2 (a), (b), (c), and (d), respectively, show the Ni $2p$, Ni $3p$, Ni$_{LMM}$ Auger transition, and O $1s$ spectra of the NiO films.
The thicknesses of the films range from submonolayer (0.5 ML) to 15.00 ML. Even visual inspection reveals monotonic shifts of the major peaks of those spectra with variation of the film thickness. However, the core-level spectra of Ni comprise many peaks of various origins, such as final state effects and non-local screening,\cite{chemical} making identification of the main peak uncertain. This complication is also transferred to the Ni$_{LMM}$ Auger transition.
Thus, we estimate shifts of the peak positions of Ni $2p$, Ni $3p$, and the Ni$_{LMM}$ Auger transition with variation in the film thickness in terms of shifts of the centroids of their spectra, 
anticipating that if shifts of the peak positions are mainly caused by extra-atomic relaxation, then all of the component peaks should shift by the same amount. 
Each spectrum is fitted with a minimal number (three or four) of Gaussian-convoluted Lorentzian peaks with Shirley backgrounds, which are used to obtain the centroid position.
The red dotted lines overlapping the experimental spectra in Fig. 2 are best-fit curves.

To determine the peak positions of the O $1s$ spectra, we fit each spectrum with Gaussian-convoluted Lorentzian peaks.
The full widths at half-maxima of major peaks are $\sim$ 2.0 eV.
The tick marks in the spectra indicate the resulting peak positions of the O $1s$ spectra. 
The curve-fitting results suggest the existence of some chemical defects, possibly Ni$_{2}$O$_{3}$, Ni(OH)$_{2}$ and/or NiO(OH), 
which appear as small shoulders in the spectra.\cite{chemical}

\begin{figure}
\centering
\includegraphics[width=0.50\textwidth]{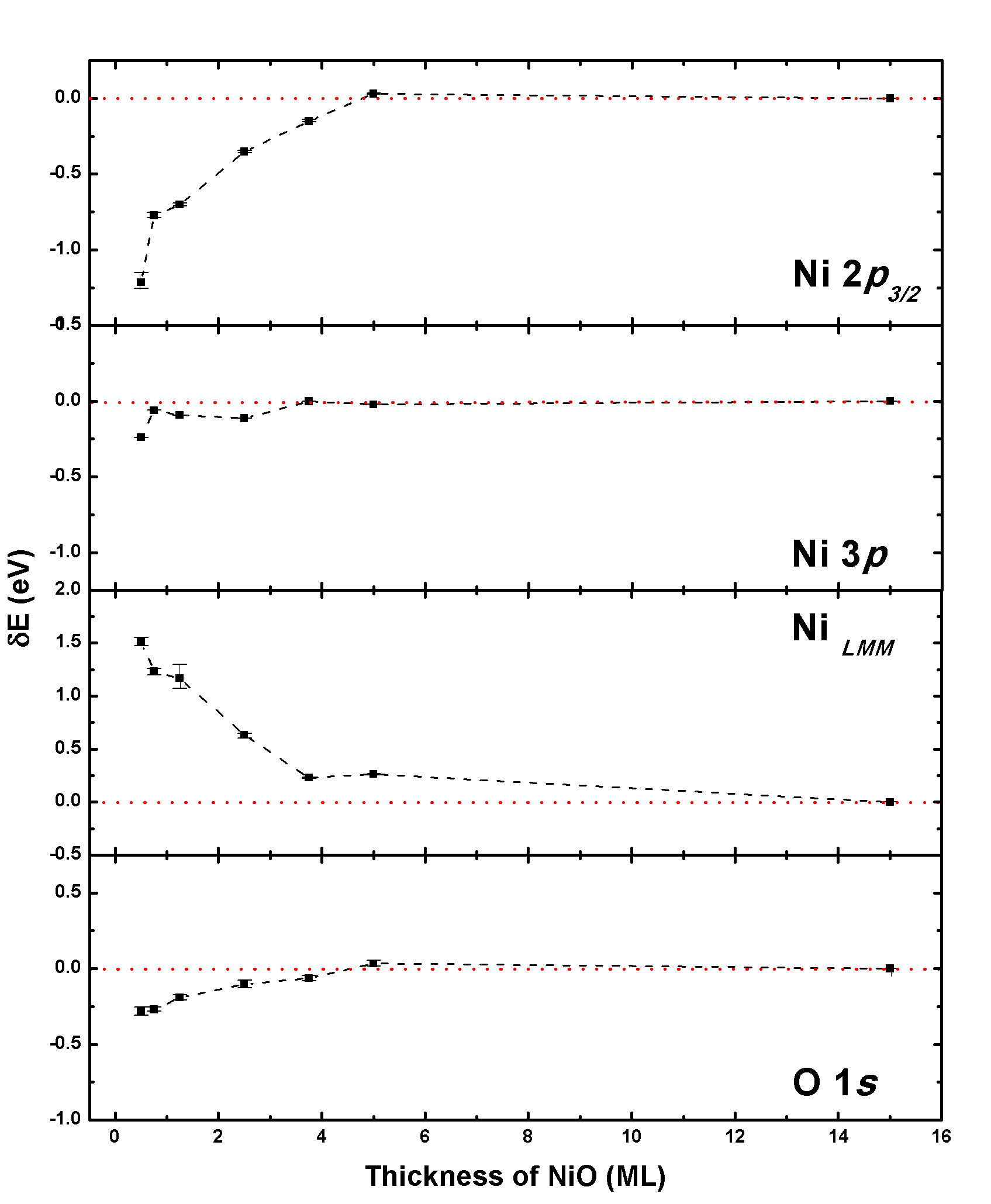}
\caption[] {The centroids of (a) Ni $2p_{3/2}$, (b) Ni $3p$, (c) Ni$_{LMM}$, and (d) the main peak position for O $1s$ relative to those of the 15 ML film are shown
as functions of the thickness of the NiO film on Ag(001). Error limits were set by the scatter of the centroid positions, depending on the fitting parameters.}
\label{fig.3}
\end{figure}

Figure 3 shows the centroid energies of Ni $2p$, $3p$, the Auger Ni$_{LMM}$ spectra, and the energy of the O $1s$ main peak as functions of the film thickness relative to the corresponding energies of the 15 ML film that is considered as a bulk-like film.
Even though the data points exhibit some scatter, we may readily observe that as the film becomes thinner, the peak positions of all the photoelectron spectra tend to shift toward the lower binding energy side, whereas the Auger transition energy of Ni$_{LMM}$ increases monotonically. For an ultrathin MgO film on Ag(001), a similar reduction of the binding energies of the photoelectrons and increase of the Auger electron energy for relevant transitions are also observed as the thickness of the film decreases.\cite{Altieri}

In Fig. 4 (a), the values of $\delta U$(Ni $3p$) obtained from Eq. (2) are plotted relative to the value for the 15 ML film.
$\delta U$(Ni $3p$) decreases monotonically as the film becomes thin, as is the case for $\delta U$(Mg $2p$)
of ultrathin MgO films on Ag(001). However, $\delta U$(Ni $3p$) changes very rapidly with increasing film thickness and is already negligible for films thicker than 5 ML.
This behavior is in contrast to that of MgO films on Ag(001)\cite{Altieri}, which exhibit substantial $\delta U$ even for 10 ML
(although both MgO and NiO of films have similar $\delta U$ values for 1 ML coverage, as shown in Fig. 4 (a)).
This is attributed to the larger polarizabilities $\alpha$(O$^{2-}$) and $\alpha$(Ni$^{2+}$) of NiO compared with MgO.
In the bulk state, $\alpha$(O$^{2-}$) of NiO (1.98 {\AA}$^{3}$) is larger than that of MgO (1.65 {\AA}$^{3}$).
Furthermore, $\alpha$(Ni$^{2+}$) $\sim$ 0.68 {\AA}$^{3}$, which is much larger than $\alpha$(Mg$^{2+}$) $\sim$ 0.09 {\AA}$^{3}$ for MgO,
possibly due to its closed-shell nature.
The larger polarizabilities of NiO should make the screening of extra charges in the cation more effective,
so extra-atomic relaxation should be more localized in NiO films than in MgO films.
Hence, in response to charge fluctuation, NiO films exhibit bulk-like behavior at smaller thicknesses than MgO films.
Note that for Fe$_2$O$_3$, $\alpha$(O$^{2-}$)$_{bulk}$ is
2 $\sim$ 2.91 {\AA}$^{3}$ (Ref. \cite{Raymond}), which is even larger than that of NiO.
Thus, for the oxide film, the coverage at which nonzero $\delta U$ is observed would be further limited according to the above argument, possibly below the experimental limit, as Chambers and Droubay\cite{Chambers} did not observe any $\delta U$ for ultrathin Fe$_2$O$_3$  films on Pt(111). These authors also attributed the absence of $\delta U$ to the large polarizabilities of the oxide.

\begin{figure}
\includegraphics[width=0.50\textwidth]{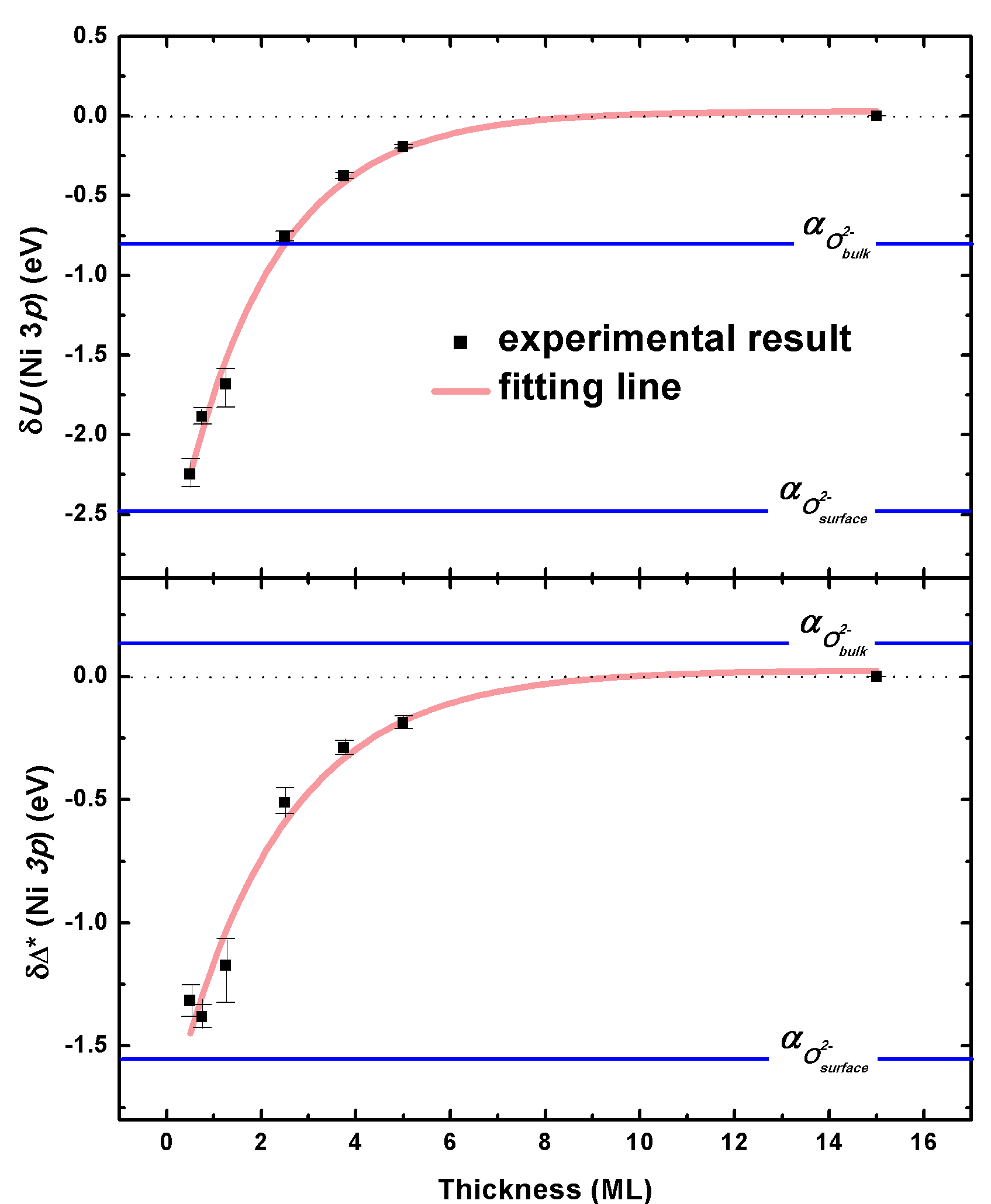}
\caption[] {Dependence of (a) $\delta U$(Ni 3p) and (b) $\delta \Delta^{\ast}$ (defined in Eq. 4)  on the nominal thickness of the NiO film on Ag(001).
The error limits are set by the fitting uncertainty. The red line in each figure is the best-fit line for the data. The blue lines show the theoretical values of $\delta U$ and $\delta \Delta^{*}$ using the respective polarizabilities of bulk and surface.}
\label{fig.4}
\end{figure}

The shifts of peak positions summarized in Fig. 3 can be suspected to originate from band bending due to charge transfer at the interface between the NiO film and Ag substrate.
However, the amount of peak shift varies for different transitions in the same film, as Fig. 3 indicates.
Hence, the peak shifts cannot be attributed to band-bending effects.
Furthermore, hybridization between the NiO film and the Ag substrate at the interface is shown to be very weak by photoelectron spectroscopy of the valence bands of the films\cite{electronic}, and this is also predicted by first principle calculations.\cite{Casassa}

\section{Discussion}


\begin{small}
\begin{table}
  \begin{center}
  \caption{Both experimental and theoretical values of $\delta U$(Ni $3p$), $\delta \Delta^{\ast}$, and E$_{image}$ of a 1 ML NiO film are summarized.
  E$_{pol}$ for bulk (1 ML film) is calculated using the polarizability of bulk (surface) NiO.
  The definition of $\delta \Delta^{\ast}$(Ni $3p$) is given in the text.}
\begin{tabular}{cc|cc}
  \hline
 \multicolumn{2}{c||}{$\alpha(O^{2-})$ ({\AA}$^{3}$)} & \multicolumn{1}{c|}{Bulk: 1.98 (Ref. \cite{Iguchi, Nakatsugawa,Moriceau})} &   \multicolumn{1}{c}{Surface: 2.43 (Ref. \cite{Iguchi, Nakatsugawa,Welton})} \\
  \hline
 \hline
 \multicolumn{2}{c||}{$\delta U_{exp}$(Ni $3p$) (eV)} & \multicolumn{2}{c}{-2.2 } \\
  \hline
 \multicolumn{2}{c||}{$\delta \Delta^{\ast}_{exp}$(Ni $3p$) (eV)} & \multicolumn{2}{c}{-1.3 } \\
  \hline
 \multicolumn{2}{c||}{$E_{image}$ (eV)} & \multicolumn{2}{c}{6.50} \\
  \hline
 \multicolumn{1}{c|}{$E_{pol}$ (eV))} & \multicolumn{1}{c||}{Bulk} &  \multicolumn{1}{c|}{-12.20}  &  \multicolumn{1}{c}{$-$} \\
    \hline
\multicolumn{1}{c|}{} & \multicolumn{1}{c||}{1 ML} &  \multicolumn{1}{c|}{$-$} &  \multicolumn{1}{c}{-8.18}  \\
    \hline
 \multicolumn{2}{c||}{$\delta U_{theo}$(Ni $3p$) (eV)} & \multicolumn{1}{c|}{$-$}  & \multicolumn{1}{c}{-2.48}  \\
  \hline
\multicolumn{2}{c||}{$\delta \Delta^{\ast}_{theo}$(Ni $3p$) (eV)} & \multicolumn{1}{c|}{$-$}& \multicolumn{1}{c}{-1.55}  \\
  \hline
\end{tabular}
  \end{center}
\end{table}
\end{small}

We investigate whether the extra-atomic relaxation represented by both $E_{image}$ and $E_{pol}$ can account
for the reduction of $U$ for the NiO films, as well as for the 1 ML MgO film on Ag(001).\cite{Altieri}
Because it is not easy to acquire layer-resolved $\delta U$ values experimentally for films thicker than 2 ML,
we calculated $E_{image}$, $E_{pol}$, and thus $\delta U$ only for a 1 ML film.
$\delta U$ is obtained from the following relation\cite{Altieri},

\begin{equation}
 \delta U = -2(E_{image} - \delta E_{pol}),
\label{eqn:3}
\end{equation}
where $\delta E_{image}$ and $\delta E_{pol}$ are both chosen to be positive,
following the convention of Altieri {\it et al.}\cite{Altieri}

The contribution of $E_{image}$ to $U$(Ni $3p$) is obtained by comparing a two-hole state, Ni $3p^{4}$, with two one-hole states, Ni $3p^{5}$.
Hence, $E_{image}$ is the difference between (2e)$^{2}/(4\pi \epsilon_{0} \times 2D$) for $3p^4$
and two one-hole states 2$\times$e$^{2}/(4\pi \epsilon_{0} \times 2D$), corresponding to $3p^5$.
Here, $D$ is the distance between a real charge and its image in the Ag substrate.
By the analysis of image potential surface states on clean Ag(001), the image plane is located 1.26 {\AA} above the Ag atoms 
in the surface layer.\cite{Smith} As a result, the Ni atoms are separated from the image plane by 1.11 {\AA}.\cite{Groppo}
Thus, according to Eq. (2), $E_{image}$ contributes $-6.50$ eV to $\delta U$  for a 1 ML film, assuming that E$_{image}$ is null for bulk NiO.

$E_{pol}$ is determined by the difference between the polarization energies of oxide for a two-hole state and two singly charged holes:
$\Sigma_{i}(4\pi\epsilon_{0} \alpha_{i} (2e)^{2}/2R_{i}^{4})$ -  2 $\Sigma_{i}(4\pi\epsilon_{0}  \alpha_{i} (e)^{2}/2R_{i}^{4})$.
For the calculation of bulk $E_{pol}$, we employ the polarizabilities $\alpha$(O$^{2-}$, Ni$^{2+}$) of bulk NiO, $\alpha$(O, Ni)$_{bulk}$,
while the polarizabilities of both O and Ni at the surface of bulk NiO, $\alpha$(O, Ni)$_{surface}$, are used to calculate $E_{pol}$ of the 1 ML film.
For $\alpha$(O$^{2-})_{bulk}$, three values have been reported: 1.49, (Ref. \cite{Kress}), 1.98 (Ref. \cite{Iguchi,Nakatsugawa,Moriceau}), and 2.64 (Ref. \cite{Janssen}) {\AA}$^{3}$. Among these, 1.98 {\AA}$^{3}$ is widely accepted. A value of 2.43 has been reported for $\alpha$(O$^{2-})_{surface}$ (Ref. \cite{Iguchi,Nakatsugawa}), which best fits the LEED I/V (spot intensity versus electron energy) of a bulk-terminated NiO(001) surface.\cite{Welton}
 $\alpha$(Ni$^{2+})$ is obtained from the empirical relationship $\alpha$(O$^{2-}$) + $\alpha$(Ni$^{2+})$
= 2.66 {\AA}$^{3}$. (Ref. \cite{Iguchi,Nakatsugawa}) This relationship was obtained for bulk NiO, but we tentatively assume that it holds down to a 1 ML film.
The resulting $E_{pol}$'s for both $\alpha$(O$^{2-})_{bulk}$ and $\alpha$(O$^{2-})_{surface}$ are summarized in Table I, along with $E_{image}$ for 1 ML NiO film.

Using the $E_{pol}$ and $E_{image}$ values in Table I, we obtain the theoretical value of $\delta U$ for a 1 ML NiO film from Eq. (3).
The experimental value of $\delta U$ (-2.2 eV) for the nominal 0.5 ML film, which is a model system for a 1 ML film, is well reproduced by the theoretical value, -2.48 eV (See Table I and Fig. 4).
This observation suggests that $E_{pol}$ and $E_{image}$ are the  major origins of $\delta U$ for NiO films, as well as for MgO films\cite{Altieri}, and reinforces the model of Duffy {\it et al.}\cite{Duffy} 
and Altieri {\it et al.}\cite{Altieri}


Altieri {\it et al.} suggested that manipulation of charge fluctuation energies of ultrathin oxide films can be used to
control their physical properties, such as N\'eel temperature $T_{N}$.\cite{Altieri}
Reduced charge fluctuation energies affect the superexchange interaction in NiO films.
According to Anderson's expression for the superexchange, the coupling constant $J$ depends on both $U$(Ni $3d$) and $\Delta$(Ni $3d$) as follows:
$J = -2t^{4}/ \Delta^{2} \times (1/\Delta + 1/U)$.
Therefore, one can expect that the reduced values of $U$(Ni $3d$) and $\Delta$(Ni $3d$) for an NiO film would lead to an increase
in the superexchange interaction for the film.
In line with this conjecture, Altieri {\it et al.} found that for a 3 ML NiO film
on Ag(001), $T_{N}$ 
did not decay as much as for an NiO film on an MgO substrate (in which no image charge screening is expected, and which therefore exhibits less
reduction of $U$ and $\Delta$).\cite{Tjeng}

To evaluate $J_{3ML}$ from Eq. (4), we use $\delta U$ (Ni 3p) in place of $\delta U$ (Ni 3d) (which is not available). 
The variation in $U$ has an extra-atomic origin, and thus we can expect that $\delta U$(Ni $3d$) will differ little
from $\delta U$(Ni $3p$). We may then estimate $\delta \Delta$ along the lines of Altieri {\it et al.}\cite {Altieri}:

\begin{equation}
 \begin{split}
 \delta \Delta (O\;\;2p \rightarrow Ni\;\;3d) & = \delta E^{bind}(O\;\;1s) - \delta E^{bind}(Ni\;\;3d) + \delta U(Ni\;\;3d)\\
                               & \approx \delta E^{bind}(O\;\;1s) - \delta E^{bind}(Ni\;\;3p) + \delta U(Ni\;\;3p)\\
                               & = \delta \Delta^{\ast}.
 \end{split}
\label{eqn.5}
\end{equation}
Here, we use $\delta U$(Ni $3p$) in place of $\delta U$(Ni $3d$) and denote the resulting $\delta \Delta$ by $\delta \Delta^{\ast}$.
From the intrapolation of $\delta U$(Ni $3p$) and $\delta \Delta^{\ast}$ in Fig. 4, the values of $\delta U$(Ni $3p$)
and $\delta \Delta^{\ast}$ for a 3 ML film are estimated to be -0.62 and -0.47, respectively.
For $U$ and $\Delta$ of bulk NiO, we use 6.5 and 4.0 eV, respectively, referring to the report of Taguchi {\it et al.}\cite{Taguchi}
Combining the above input, $J$ is found to be $-2t^{2}$ $\times$ 0.0364 for a 3 ML NiO film on Ag, while $J$ for bulk NiO is
$-2t^{2}$ $\times$ 0.0252. Here, we assume the transfer integral $t$ between the anion O $2p$ and the cation Ni $3d$ is the same for both bulk and the 3 ML film, as $t$ is a very local property and is assumed to be little influenced by extra-atomic effects.

We can now estimate $T_{N}$ for a 3 ML film in the mean field approximation. 
In the mean field approximation, $T_{N} \sim S(S + 1) \times N \times J$, where S is the spin moment of an Ni ion, and N the mean number of nearest neighbors of Ni ions. Then,

\begin{equation}
 T_{N,film} = T_{N,bulk}  \times S_{film}(S_{film} + 1)/S_{bulk}(S_{bulk} + 1) \times (N_{film}/N_{bulk}) \times (J_{3ML}/J_{bulk})
\label{eqn:6}
\end{equation}
For a bulk-like thick NiO film on Ag(001), $T_{N}$ was experimentally determined to be 535 K.\cite{Tjeng} 1.90 (Ref. \cite{Cheetham}) and 2.2 (Ref. \cite{Fernandez, Neubeck}) $\mu_B$ have been reported for the total magnetic
moment $M_{bulk}$ of bulk NiO. First principle calculations
predict that $M_{3ML}$ of a 3 ML NiO film on Ag(001) reduces to $\sim$ 1.67 $\mu_{B}$ (the average of the moments of the 1st, 2nd, and
3rd layers).\cite{Cinquini}  If the ratio $L/S$ of the orbital moment to the spin moment is assumed to be the same (0.34)
(Ref. \cite{Fernandez, Neubeck}) for both bulk
and film, $S_{bulk}$ is 0.81 (Ref. \cite{Cheetham}) or 0.94 (Ref. \cite{Fernandez, Neubeck}), and $S_{3ML}$ is 0.71. The mean number of nearest neighbors of a 3 ML film is $\sim$ 9.33. If all this input is taken into account, then according to Eq. (6),
$T_{N}$ is between 400 and 498 K for a 3 ML NiO film on Ag(001).
The wide variation in $T_{N}$ originates mainly from the large uncertainty in the spin moment of bulk NiO. The
experimentally determined $T_{N}$ of a 3 ML NiO film is 390 K\cite{Tjeng}, which  is  close to the range of the present estimate. Despite the many
simplifications and assumptions, $\delta U$(Ni $3p$) and $\delta \Delta^{*}$(Ni $3p$) seem to provide a reasonable estimate of the range of $T_{N}$ for
an NiO film under the mean field approximation.
Most importantly, the lower limit of the present estimate (400.00 K) is still much higher than the $T_{N} \sim$ 40 K observed for a 3 ML NiO film on an MgO substrate.\cite{Tjeng}
At the very least, this supports the argument that the reduction of charge fluctuation energies on a polarizable substrate gives rise to high values of $T_{N}$ (as observed for the 3 ML NiO film on Ag(001) in comparison with the results for MgO(001)).

\section{Summary and Conclusion}

Using both photoelectron spectroscopy and Auger electron spectroscopy, we found that the on-site Coulomb interaction energy $U$ of ultrathin NiO films on Ag(001) decreases monotonically in a manner analogous to the case of ultrathin MgO films on Ag(001). The observed value of $\delta U$ (Ni $3p$) for a 1 ML film was well reproduced by considering extra-atomic relaxations represented by image charge screening by the Ag substrate and modified polarization energy of the film, thus affirming the pictures of Duffy {\it et al.}\cite{Duffy} and Altieri {\it et al.}\cite{Altieri} Furthermore, using $\delta U$ (Ni $3p$), we estimated the value of T$_N$ for a 3 ML NiO film on Ag(001), and the estimate was comparable to experimental observation. Hence, the model proposed by the aforementioned authors seems to offer a unified picture of the variation in charge fluctuation energies of ultrathin MgO and NiO films, even though further refinement is necessary in view of the many assumptions/estimations employed without precise quantitative justification.

\section{Acknowledgement}
This work was supported by the KOSEF grant No. R01-2007-000-20249-0 and the NRF grant No. 20110004239.

\end{document}